\documentstyle[floats,preprint,prl,aps]{revtex}

\tightenlines
\begin{document}

\draft
\title{Noise-Activated Escape from a Sloshing Potential Well}
\author{Robert S. Maier${}^{1,2}$ and D.~L. Stein${}^{2,1}$}
\address{${}^{1}$Mathematics and ${}^{2}$Physics Departments, University of
Arizona, Tucson, Arizona 85721}

\maketitle

\begin{abstract}
We treat the noise-activated escape from a one-dimensional potential well
of an overdamped particle, to which a periodic force of fixed frequency is
applied.  We determine the boundary layer behavior, and the physically
relevant length scales, near the oscillating well top.  We show how
stochastic behavior near the well top generalizes the behavior first
determined by Kramers, in the case without forcing.  Both the case when the
forcing dies away in the weak noise limit, and the case when it does not,
are examined.  We also discuss the relevance of various scaling regimes to
recent optical trap experiments.
\end{abstract}


\narrowtext

The phenomenon of weak white noise inducing escape from a one-dimensional
potential well was studied by Kramers~\cite{Kramers40}.  If~$\epsilon$
denotes the noise strength (e.g., $\epsilon\propto k_BT$ in thermal
systems), and $\Delta E$ measures the depth of the well, then the escape
rate~$\lambda$ falls~off like $\exp(-\Delta E/\epsilon)$ as~$\epsilon\to0$.
The case when the trapped particle is overdamped is easiest to analyse.  If
the particle, after each escape, is reinjected at the bottom of the well,
and a steady state has been set~up, then in the interior of the well its
position will have a Maxwell--Boltzmann distribution.  Kramers determined
that this distribution must be modified near the well top, by being
multiplied by a `boundary layer function' that incorporates outgoing
boundary conditions.  From this modified distribution, he was able to
determine the weak-noise limit of the escape rate, including the
all-important pre-exponential factor, by computing the probability flux
over the well top.

The Kramers formula and its multidimensional generalization have been
extended in many ways~\cite{McClintock89a,Talkner95}.  There have been
extensions to non-overdamped particles and to colored noise.  There have
also been extensions to the case when even though the particle is
overdamped and the noise is white, the noise-perturbed dynamics of the
particle fail to satisfy detailed balance.  This may be due to localized
`hot spots'~\cite{Landauer83} or, in multidimensional systems, to
nonconservative deterministic dynamics~\cite{MaierB}.

However, there is one experimentally important case that has not been
exhaustively studied.  That is when the system parameters are {\em
periodically modulated\/}.  A~full analysis of escape driven by weak noise,
in such systems, would shed light on the Kramers limit of stochastic
resonance.  It~would also clarify the effects of barrier modulation on
phase-transition phenomena.

It is now possible to construct a physical system (a~mesoscopic dielectric
particle that moves, in an overdamped way, within a dual optical
trap~\cite{McCann99}) that provides a clean experimental test of the
three-dimensional Kramers formula.  The rate at which thermal noise induces
escape agrees well with the predictions of the formula.  Adding an external
force, of fixed period~$\tau_F$, would yield a periodically modulated
system~\cite{Simon92}, of the sort that has not yet been fully analysed.
A~complete treatment of escape from a well of a periodically driven
overdamped particle, or equivalently the escape of an overdamped particle
from a `sloshing potential well', would be desirable.

Smelyanskiy, Dykman, and Golding treated this phenomenon perturbatively, in
one dimension~\cite{Smelyanskiy99}.  They derived a Kramers prefactor
incorporating~$f$, the strength of the periodic forcing.  It~applies if the
ratio $f/\epsilon$ is set to a constant as~$\epsilon\to0$.  That~is, the
forcing is taken to die away in the weak-noise limit.  Lehmann, Reimann,
and H{\"a}nggi~\cite{Lehmann00} treated nonperturbatively the case when
$f$~is independent of~$\epsilon$, using path integral techniques, and
worked~out a numerical scheme for computing the $f$-dependent prefactor.
They also examined the `instantaneous escape rate', which in the steady
state is a $\tau_F$-periodic function of time.  In a simulation of a
special case (a~well with a perfectly harmonic top), they noted that in the
weak-noise limit, the maximum of the instantaneous escape rate cycles
slowly around the interval $[0,\tau_F)$.

In this Letter, we go beyond \cite{Smelyanskiy99} and~\cite{Lehmann00}.  By
treating the case $f\propto\epsilon^\alpha$, where $\alpha$~is an arbitrary
nonnegative power of~$\epsilon$, we determine the relation between their
respective scaling regimes.  In the weak-noise, weak-forcing limit, there
are three physically important length scales near the oscillating well top,
of sizes proportional to~$\epsilon^{1/2}$, $f$, and $f^{1/2}$.  Crossover
behavior will result if $f\propto\epsilon^{1/2}$, and the case
$f\propto\epsilon$ can itself be viewed as a crossover regime.

When $f$~is independent of~$\epsilon$, we use facts on noise-induced
transport through unstable limit cycles to illuminate the `cycling'
phenomenon~\cite{Day93}.  At any~$t$ in $[0,\tau_F)$, the normalized
instantaneous escape rate oscillates periodically in $\log\epsilon$
as~$\epsilon\to0$.  We~supply a formula for the period, and give a physical
explanation for the logarithmic slowness.

More importantly, we place the case of $\epsilon$-independent periodic
forcing firmly in the Kramers framework, by determining how the
Maxwell--Boltzmann distribution is modified, in the boundary layer of width
${\cal O}(\epsilon^{1/2})$ near the oscillating well top.  As~$f\to0$, it
approaches the modified distribution of Kramers~\cite{Kramers40}.  The case
when $f\propto\epsilon$ in the weak-noise limit is intermediate between the
case of $\epsilon$-independent forcing and the case of zero forcing, and
its boundary layer behavior is intermediate too.

{\em Scaling Regimes.}---Initially, we work in~terms of dimensional
quantities.  The Langevin equation for a driven Brownian particle in a
potential well $U=U(x)$ is
\begin{equation}
m\ddot x + \gamma m \dot x = -U'(x) + F\nu(t) + \sqrt{2m\gamma k_B T}\, \eta(t).
\end{equation}
Here $\gamma$ is the damping, $F$~a dimensional measure of the driving,
$\nu$~a dimensionless periodic function of unit amplitude, and $\eta$ a
standard white noise.  In~the overdamped (large-$\gamma$) limit, the
inertial term can be dropped, leaving
\begin{equation}
\label{eq:Langevin}
\dot x = -V'(x) + f\nu(t) + \sqrt\epsilon \,\eta(t).
\end{equation}
Here $V=U/\gamma m$, $f=F/\gamma m$, and $\epsilon=2k_B T/\gamma m$.  

The Kramers formula for the $f=0$ escape rate is
\begin{eqnarray}
\lambda&\sim&\frac{\omega_s\omega_u}{2\pi\gamma} \exp{(-\Delta U/k_BT)}
\nonumber\\
&=&\frac{\sqrt{V''(x_s)\left|V''(x_u)\right|}}{2\pi} \exp{(-\Delta
E/\epsilon)},
\label{eq:Kramers}
\end{eqnarray}
where $\omega_s = \sqrt{U''(x_s)/m}$ and $\omega_u=
\sqrt{\left|U''(x_u)\right|/m}$ are the oscillation frequencies about the
bottom~$x_s$ and top~$x_u$ of the well, and $\Delta E=2\Delta V$.
Eq.~(\ref{eq:Kramers}) follows from Kramers's modification of the
steady-state Maxwell--Boltzmann weighting $\exp\left[-U(x)/k_BT\right]$,
i.e., $\exp\left[-2V(x)/\epsilon\right]$.  If $n$ denotes the inward offset
from~$x_u$, his modifying factor is ${\rm
erfc}\bigl[-n/\sqrt{\epsilon/|V''(x_u)|}\bigr]$.

If $f\neq0$, there are two regimes, depending on the size of~$f$
as~$\epsilon\to0$.  Since $[\epsilon] =[\Delta E]=L^2/t$ and $[f]=L/t$,
where $L$~denotes length and $t$~denotes time, comparing $f$
with~$\epsilon$ must be done with care.  $f$~will be `small' or `large' in
the Kramers limit if it is small or large compared to a quantity with
dimensions~$L/t$, namely $\sqrt{\left|V''(x_u)\right|\epsilon}$.

In physical terms, there are two regimes because there are two length
scales at the well top, and one or the other is larger.  The first is the
length scale in Kramers's modification.  There is a boundary layer of width
$\approx\sqrt{2k_BT/\left|U''(x_u)\right|}$, i.e.,
$\sqrt{\epsilon/|V''(x_u)|}$, within which `physics occurs'.  This ${\cal
O}(\epsilon^{1/2})$ quantity is the diffusion length: the distance from the
top to which the particle must approach, to acquire a substantial chance of
leaving the well rather than falling back.

If a periodic force is applied, a second length scale becomes important.
The top of the well will oscillate periodically around the unperturbed
top~$x_u$ by an amount roughly equal to $F/\left|U''(x_u)\right|$.  If~this
length scale is substantially smaller than the first, to a first
approximation the boundary layer will not oscillate.  But if the opposite
is true, escape dynamics should be strongly affected by boundary layer
oscillations.  The crossover occurs when
$F\approx\sqrt{2\left|U''(x_u)\right|k_BT}$, i.e., when
$F\approx\sqrt{2m\omega_u^2k_BT}$.  In normalized units, this criterion is
$f\approx \sqrt{\left|V''(x_u)\right|\epsilon}$.

So if $f\propto\epsilon^\alpha$ in the Kramers limit ($\epsilon\to0$),
$\alpha>1/2$ and ${0\le\alpha<1/2}$ belong to different regimes.  The
$\alpha=1$ results of Ref.~\cite{Smelyanskiy99} presumably extend to the
entire $\alpha>1/2$ regime.  Similarly, our treatment of $\alpha=0$ below
could be extended to cover the ${0\le\alpha<1/2}$ regime.  These two
regimes should be kept in mind when conducting experiments on noise-driven
escape in periodically driven systems.  In~the Kramers limit, only when the
forcing~$F$ is much less than $\sqrt{2m\omega_u^2k_BT}$ is a simple
perturbative modification of the Kramers formula likely to apply.

An illustration would be the room-temperature dual optical trap experiment
of McCann, Dykman, and Golding~\cite{McCann99}, in which $m\approx
3\times10^{-16}\,{\rm kg}$ and $\omega_u={(7\pm2)}\times10^4\,{\rm
sec}^{-1}$.  The corresponding force magnitude $\sqrt{2m\omega_u^2k_BT}$ is
approximately $10^{-13}$ Newtons.  Any repetition of their experiment, with
the addition of periodic driving, should take this dividing line into
account.

Another scaling-related issue has to do with the effects of choosing a
period $\tau_F$ for the forcing that is very small or large.  In~the
Kramers limit, it is possible to take $\tau_F\propto\epsilon^\beta$, where
$\beta$ may be positive or negative.  We~make the natural choice $\beta=0$,
so that $\tau_F$ is independent of~$\epsilon$.

{\em Preliminaries.}---Our analysis of the $\alpha=0$ case uses {\em
optimal trajectories\/}.  The $\epsilon\to0$ limit is governed by the
action functional
\begin{equation}
\label{eq:noforceaction}
{\cal W}\left[t\mapsto x(t)\right] = \frac12
\int \left| \dot x + V'(x) - f\nu(t)\right|^2\,dt.
\end{equation}
First, suppose that $f=0$.  Then the most probable trajectory from $x_s$ to
any specified point~$x'$ is the one that minimizes ${\cal W}\left[t\mapsto
x(t)\right]$.  The minimum is taken over all trajectories from $x_s$
to~$x'$, and all transit times (infinite as well as finite).  There is a
single minimizer $t\mapsto x_*(t)$ to each side of~$x_s$, which we term an
optimal trajectory.  The value ${\cal W}\left[t\mapsto x_*(t)\right]$,
which depends on the endpoint~$x'$ and may be denoted $W(x')$, is the rate
at which fluctuations to~$x'$ are exponentially suppressed
as~$\epsilon\to0$.  In the steady state, the probability density
$\rho=\rho(x)$ of the particle will have the asymptotic form
\begin{equation}
\label{eq:asymptoticform}
\rho(x)\sim K(x)\exp\left(-W(x)/\epsilon\right),\qquad \epsilon\to0.
\end{equation}
The prefactor $K(x)$ must be computed by other means.

Any such $f=0$ optimal trajectory must satisfy $\dot x=+V'(x)$, i.e., be a
time-reversed relaxational trajectory.  This is due to detailed balance,
which holds in the absence of `hot spots'.  The optimal trajectory from
$x_s$ to~$x_u$ is instanton-like: it~emerges from~$x_s$ at $t=-\infty$ and
approaches $x_u$ as $t\to+\infty$.  Within the well, $W(x)$ equals
$2[V(x)-V(x_s)]$, so $\Delta E\equiv W(x_u)$ equals
$2\left[V(x_u)-V(x_s)\right]$.  Also, $K$~is independent of~$x$.

If $f=0$, the model defined by the Langevin equation~(\ref{eq:Langevin}) is
invariant under time translations.  So the optimal trajectory from $x_s$
to~$x_u$ is not unique.  If $x=x_*(t)$ is a reference optimal trajectory,
consider the family
\begin{equation}
\label{eq:family}
t\mapsto x^{(\phi)}_*(t)\equiv x_*(t+\frac\phi{2\pi} \tau_F),
\end{equation}
where the phase shift~$\phi$ satisfies $0\le\phi<2\pi$, and $\tau_F$ is the
period of the forcing function $\nu=\nu(t)$.  In the Kramers limit of
any model with $f$ nonzero but small, the most probable escape trajectory
should resemble some trajectory of the form~(\ref{eq:family}).  That~is,
some $\phi_m$ will be singled~out as maximizing the chance of a particle
being `sloshed out'.  A~study of the $f\to0$ limit should yield~$\phi_m$.

This was the approach of \cite{Smelyanskiy99}.  Suppose that $f\neq0$.  If
$\Delta E$ is computed by applying (\ref{eq:noforceaction}) to the
unperturbed ($f=0$) optimal trajectory $x=x^{(\phi)}_*(t)$, the first-order
(i.e., ${\cal O}(f)$) correction to $\Delta E$ will be $fw_1(\phi)$, where
\begin{equation}
\label{eq:w1}
w_1(\phi) \equiv - \int_{-\infty}^\infty \dot x^{(\phi)}_*(t)\nu(t)\,dt.
\end{equation}
It is reasonable to average the Arrhenius factor $\exp\left(-\Delta
E/\epsilon\right)$ in the Kramers formula over~$\phi$, from $0$ to~$2\pi$.
If $\langle\bullet\rangle_\phi$ denotes this averaging, then the escape
rate will be modified by the driving, to leading order, by a factor
$\langle e^{-fw_1(\phi)/\epsilon}\rangle_\phi$.  If $\alpha=1$, i.e.,
$f=f_1\epsilon$ for some~$f_1$, then the Kramers formula~(\ref{eq:Kramers})
will be altered to
\begin{equation}
\label{eq:altered}
\lambda\sim
\langle e^{-f_1w_1(\phi)}\rangle_\phi
\frac{\sqrt{V''(x_s)\left|V''(x_u)\right|}}{2\pi} \exp{(-\Delta
E/\epsilon)}.
\end{equation}
Clearly, $\phi_m$ should be the phase that minimizes $w_1(\phi)$.

Eq.~(\ref{eq:altered}) is essentially the formula of Smelyanskiy et
al.~\cite{Smelyanskiy99}.  But our derivation makes it clear that their
perturbative approach requires that $f\to0$ rapidly as~$\epsilon\to0$,
i.e., that $\alpha$~be sufficiently large.  Estimating the minimum
of~${\cal W}[\bullet]$ by applying it to {\em unperturbed\/} optimal
trajectories yields a correction to~$\Delta E$ which is valid only
to~${\cal O}(f^1)$.

If $f$ is independent of~$\epsilon$, then by Laplace's method
\begin{equation}
\langle
e^{-fw_1(\phi)/\epsilon}\rangle_\phi
\sim
\frac{1}{\sqrt{2\pi w''_1(\phi_m)f}}
\,{\epsilon}^{1/2}
e^{-fw_1(\phi_m)/\epsilon}
\end{equation}
as~$\epsilon\to0$.  This would seemingly suggest that
\begin{equation}
\label{eq:modifiedKramers}
\lambda\sim
\frac{\sqrt{V''(x_s)\left|V''(x_u)\right|}}{2\pi}
\frac{1}{\sqrt{2\pi w''_1(\phi_m)f}}
\,{\epsilon}^{1/2}
\exp{(-\Delta E/\epsilon)},
\end{equation}
where $\Delta E$ is shifted by $fw_1(\phi_m)$ to leading order, is the
$\alpha=0$ Kramers formula.  But the prefactor
in~(\ref{eq:modifiedKramers}) is correct only in the small-$f$ limit.  If
$f\propto\epsilon^{\alpha}$, the ${\cal O}(f^1)$ correction to~$\Delta E$
will be of magnitude $\epsilon^\alpha$.  If $\alpha=1$, it will induce, as
in~(\ref{eq:altered}), a correction to the prefactor.  But when
$\alpha\le1/2$, ${\cal O}(f^2)$ corrections will also affect the prefactor.
The most difficult case is $\alpha=0$, when computing the prefactor would
require working to all orders in~$f$.  A~nonperturbative treatment, like
the analysis of Lehmann et~al.~\cite{Lehmann00} or the following analysis,
is needed.

{\em Analysis.}---We first remove explicit time-dependence, when $f\neq0$
and $\tau_F$ are fixed, by replacing~(\ref{eq:Langevin}) by
\begin{eqnarray}
\nonumber
\dot x &=& -V'(x) + f\nu(y) + \sqrt\epsilon \,\eta(t),\\
\dot y &=& 1.
\label{eq:Langevin2D}
\end{eqnarray}
Here $0\le y<\tau_F$, and $y$~is periodic: $y=\tau_F$ is identified with
$y=0$.  The state space with coordinates ${\bf X}\equiv(x,y)$ is
effectively a cylinder.  On~this cylinder, the oscillating well bottom
$x=\tilde x^{(f)}_s(t)$ is a stable limit cycle, and the oscillating well
top $x=\tilde x^{(f)}_u(t)$ is an unstable limit cycle.  To~stress
$f$-dependence, we~denote them ${\bf X}_s^{(f)}$ and~${\bf X}_u^{(f)}$.

To study escape through ${\bf X}_u^{(f)}$ when ${\epsilon\to0}$, we can
employ results of Graham and T\'el~\cite{Graham84,Graham85}.  The limit is
governed by an instanton-like optimal trajectory ${\bf X}={\bf
X}_*^{(f)}(t)$ that spirals out of~${\bf X}_s^{(f)}$ and into~${\bf
X}_u^{(f)}$.  It~is the most probable escape trajectory in the steady
state.  The exponent $\Delta E$ equals ${\cal W}\bigl[t\mapsto{\bf
X}_*^{(f)}(t)\bigr]$, which in~general must be computed numerically.  The
trajectory ${\bf X}_*^{(f)}$ would be computed nonperturbatively, by
integrating Euler--Lagrange or Hamilton equations.

${\bf X}_*^{(f)}$ increasingly resembles a time-reversed relaxational
trajectory, as it nears the oscillating well top.  So at any specified~$y$,
the $l$th winding of ${\bf X}_*^{(f)}$, as it spirals into~${\bf
X}_u^{(f)}$, has an inward offset~$n$ that shrinks geometrically,
like~$ac^{-l}$, as~$l\to\infty$.  Here $a=a(y)$ and~$c$ are $f$-dependent,
and $c=\exp\bigl[\oint \bigl|V''(\tilde x_u^{(f)}(t))\bigr|\,dt\bigr]$.

The form (\ref{eq:asymptoticform}) for the steady-state probability density
generalizes to $K({\bf X})\exp\left(-W({\bf X})/\epsilon\right)$.
To~compute $W$ and~$K$ at any specified~${\bf X}'$, an optimal trajectory
ending at~${\bf X}'$ is needed; in~general, one different from~${\bf
X}_*^{(f)}$.  An asymptotic analysis of the Smoluchowski equation for the
probability density~\cite{MaierB,Maier12} shows that $W$~satisfies the
Hamilton--Jacobi equation
\begin{equation}
(\bbox{\nabla}W)\cdot {\bf D} \cdot (\bbox{\nabla} W)/2 + {\bf u}\cdot
\bbox{\nabla}W = 0,
\label{eq:HJ}
\end{equation}
and along any optimal trajectory, $K$ satisfies
\begin{equation}
\label{eq:K}
\dot K = -(\bbox{\nabla}\cdot{\bf u} + D_{ij} \partial_i\partial_j W/2) K.
\end{equation}
Here ${\bf u}(x,y)\equiv(-V'(x)+f\nu(y),1)$ is the drift on the cylinder,
and $(D_{ij})={\rm diag}(1,0)$ is the diffusion tensor.  It follows
from~(\ref{eq:HJ}) that the Hessian matrix $(\partial_i\partial_j W)$ obeys
a Riccati equation along any optimal trajectory \cite{Maier12,Ryter84}.
This gives a numerical scheme for computing $K({\bf X}')$.  By convention,
$K$~is chosen to be ${\cal O}(1)$ on~${\bf X}_s^{(f)}$.

In principle, the steady-state escape rate $\lambda$ can be computed by the
Kramers method~\cite{Kramers40}: evaluating the probability flux
through~${\bf X}_u^{(f)}$.  But this is intricate, due to a subtle problem
discovered by Graham and T\'el~\cite{Graham84,Graham85}.  Optimal
trajectories that are perturbations of the escape trajectory $t\mapsto{\bf
X}_*^{(f)}(t)$ intersect one another wildly near~${\bf X}_u^{(f)}$.  This
is because $t\mapsto{\bf X}_*^{(f)}(t)$ is a delicate object: a `saddle
connection' in the Hamiltonian dynamics sense.  In consequence, any ${\bf
X}'$ near~${\bf X}_u^{(f)}$ is reached by an {\em infinite discrete set\/}
of optimal trajectories, indexed by~$l$, the number of times a trajectory
winds around the cylinder before reaching~${\bf X}'$.  The density
asymptotics are~\cite{Maier12}
\begin{equation}
\label{eq:newnewasymptoticform}
\rho({\bf X})\sim \sum_l K^{(l)}({\bf X})\exp\bigl(-W^{(l)}({\bf X})/\epsilon\bigr),
\qquad \epsilon\to0,
\end{equation}
since $W$~and~$K$ are {\em infinite-valued\/}, not single-valued.

It~is known~\cite{Graham84,Graham85,Maier12} that at any fixed~$y$, any
$W^{(l)}$ is not quadratic but linear in the offset~$n$ from~${\bf
X}_u^{(f)}$:
\begin{equation}
\label{eq:foobar}
W^{(l)}(n)\approx \Delta E - \left|W_{,nn}\right| \left(ac^{-l}n-(a c^{-l})^2/2\right).
\end{equation}
$W_{,nn}<0$ is what, in the absence of multivaluedness, the Hessian matrix
element $\partial^2 W/\partial n^2$ would equal at ${n=0}$.  Along~${\bf
X}_u^{(f)}$, it obeys the scalar Riccati equation
\begin{equation}
\label{eq:scalarRiccati}
\partial W_{,nn}/\partial y = -{W_{,nn}}^2 
+2 V''\bigl(\tilde x^{(f)}_u(y)\bigr)W_{,nn}.
\end{equation}
$W_{,nn}=W_{,nn}(y)$ is the $\tau_F$-periodic solution of this equation,
which is easy to solve numerically.  At any~$y$, $W_{,nn}$ equals
$2V''(x_u)$ to leading order in~$f$.  Deviations from this value are due to
anharmonicity of~$V$ at the well top.

It~is also known~\cite{Maier12} that the second term on the right-hand side
of~(\ref{eq:K}) tends rapidly to zero along ${\bf X}_*^{(f)}$, as it
spirals into~${\bf X}_u^{(f)}$.  So~with each turn, $K$~is multiplied by
$\exp\bigl[-\oint({\bf \nabla}\cdot{\bf u})\,dt\bigr]$, i.e., by
$\exp\bigl[\oint V''(\tilde x_u^{(f)}(t))\,dt\bigr]$.  This factor
equals~$c^{-1}$.  So $K^{(l)}\sim Ac^{-l}$ for some~$A=A(y)$.  Since $n\sim
ac^{-l}$, it follows that along~${\bf X}_*^{(f)}$, $K\sim k_1n$ as~$n\to0$.
Here $k_1 \equiv A/a$, like~$W_{,nn}$, is a $\tau_F$-periodic function
of~$y$, which quantifies the linear falloff of~$K$ near ${\bf X}_u^{(f)}$.
The linear falloff of~$K$ is a nonperturbative effect.

As a function on~$[0,\tau_F)$, $k_1$ turns~out to be proportional
to~$W_{,nn}$~\cite{MaierSteininpreparation}.  It~can be obtained
numerically by integrating~(\ref{eq:K}) along the trajectory~${\bf
X}_*^{(f)}$, as it spirals into~${\bf X}_u^{(f)}$.  It~is the $t\to\infty$
limit of the quotient $K/n$.  Deviations from constancy are due to
anharmonicity of~$V$.

Substituting (\ref{eq:foobar}) and $K^{(l)}\sim k_1ac^{-l}$
into~(\ref{eq:newnewasymptoticform}) yields 
\begin{equation}
e^{-\Delta E/\epsilon} \sum_{l=-\infty}^\infty k_1 ac^{-l} \exp \left\{ \left|W_{,nn}\right|\left[ac^{-l}n -
(ac^{-l})^2/2\right]/\epsilon \right\}
\label{eq:fundamental}
\end{equation}
as the $\epsilon\to0$ steady-state probability density~$\rho$, at an inward
offset~$n$ from the oscillating well top.  Summing from $-\infty$
to~$\infty$ is acceptable since the errors it introduces are exponentially
small, and can be ignored.  The dependence here on~$t$, i.e., on~$y$, is
due to $W_{,nn}$, $k_1$, and~$a$.

{\em Discussion.}---The cycling phenomenon, and much else, follow from the
infinite sum~(\ref{eq:fundamental}).  To determine its behavior on the
${\cal O}(\epsilon^{1/2})$ diffusive length scale near the oscillating well
top, set $n=N\epsilon^{1/2}$ with $N$~fixed, and also multiply
by~$\epsilon^{-1/2}$.  (As~in the case of no periodic driving, a
steady-state density~$\tilde\rho$ that is normalized to total
probability~$1$ within the well must include an $\epsilon^{-1/2}$ factor.)
The resulting expression is invariant under $\epsilon\mapsto
c^{-2}\epsilon$.  So
\begin{equation}
\tilde\rho (n=N\epsilon^{1/2},t) \sim h_\epsilon^{(f)} (N,t) \exp (-\Delta
E/\epsilon),\quad \epsilon\to0,
\label{eq:last}
\end{equation}
where the quantity $h_\epsilon^{(f)}(N,t)$, for any~$N$ and any $t$
in~$[0,\tau_F)$, is periodic in $\log\epsilon$ with period $2\log c$.

In the steady state, the instantaneous escape rate $\lambda(t)$ through the
oscillating well top, which equals $(\epsilon/2)(\partial/\partial
n)\tilde\rho\,|_{n=0}$, satisfies
\begin{equation}
\lambda(t) \sim (1/2)\epsilon^{1/2} h_\epsilon^{(f)\prime}(0,t)\exp (-\Delta
E/\epsilon), \qquad \epsilon\to0.
\end{equation}
So at any $t$ in~$[0,\tau_F)$, the instantaneous escape rate, divided
by~$\epsilon^{1/2}$, ultimately oscillates periodically in $\log\epsilon$
with period $2\log c$, i.e., with period $2\bigl[\oint \bigl|V''(\tilde
x_u^{(f)}(t))\bigr|\,dt\bigr]$.

Lehmann et~al.~\cite{Lehmann00} noticed that on~$[0,\tau_F)$, the peak of
the function~$\lambda(\bullet)$ may shift when $\epsilon$ is decreased.
Our results indicate that slow oscillations in the instantaneous escape
rate are a widespread phenomenon.  They have a simple physical cause.  In
the $\epsilon\to0$ limit, the most probable trajectory taken by an escaping
particle is the helix $t\mapsto{\bf X}_*^{(f)}(t)$, along which it moves in
a ballistic, noise-driven way.  However, once it gets within an ${\cal
O}(\epsilon^{1/2})$ distance of the oscillating well top, it moves
diffusively rather than ballistically.  It~is easily checked that the
changeover to diffusive behavior takes place at a location that cycles
slowly around $[0,\tau_F)$, as~$\epsilon\to0$.  If $\epsilon\mapsto
c^{-2}\epsilon$, the changeover returns to its original location.

If the well top is perfectly harmonic, so that $W_{,nn}$ and $k_1$ do not
depend on~$t$, and the bottom is too, it is straightforward to integrate
$\lambda(t)$ over $[0,\tau_F)$.  We~find
\begin{equation}
\label{eq:trueKramers}
\lambda \sim
\frac{k_1 \sqrt{V''(x_s)}}
{\sqrt{2\pi}\,\tau_F\left|V''(x_u)\right|}
\epsilon^{1/2}\exp{(-\Delta E/\epsilon)}.
\label{eq:special}
\end{equation}
It is useful to compare (\ref{eq:trueKramers}) with the perturbative
formula~(\ref{eq:modifiedKramers}).  They can be reconciled if $k_1$
diverges like~$f^{-1/2}$ as~${f\to0}$.  An~$f^{-1/2}$ divergence was seen
in this special case by Lehmann et~al.~\cite{Lehmann00}, and it occurs more
widely~\cite{MaierSteininpreparation}.  It has major consequences.
$k_1$~is the normal derivative of the density prefactor~$K$.  But $K$~is
${\cal O}(1)$ on~${\bf X}_s^{(f)}$, and is well-behaved in the well
interior as~$f\to0$.  So there must be a layer near the well top, of width
${\cal O}(f^{1/2})$, within which $K$ slopes~off to zero.  The presence of
this layer has been numerically confirmed~\cite{MaierSteininpreparation}.

We can now compare the steady-state probability density~(\ref{eq:last}),
which is valid on the ${\cal O}(\epsilon^{1/2})$ length scale near the
oscillating well top, to the density when $f=0$ on the same length scale.
The analog of $h_\epsilon^{(f)}(N,t)$, if $f=0$, is
\begin{displaymath}
{\rm
erfc}\left[-\sqrt{|V''(x_u)|}\,N\right] \times
\epsilon^{-1/2}\exp\left(\left|V''(x_u)\right|N^2\right),
\end{displaymath}
up~to a constant.  The first factor is the Kramers boundary layer
function~\cite{Kramers40}, and the second is from the Maxwell--Boltzmann
distribution.

It may seem odd that $h_\epsilon^{(f)}(N,t)$, which is defined by a
complicated infinite sum, should degenerate into such a classical (and
$t$-independent) form in the $f\to0$ limit.  The details remain to be
worked~out, but the mechanism is clear: the $f\to0$ limit passes through an
intermediate scaling regime, namely $\alpha=1$, where
(\ref{eq:fundamental})~does not apply.  The dominant terms in the
sum~(\ref{eq:fundamental}) are those for which $ac^{-l}$ is comparable
to~$\epsilon^{1/2}$.  But in deriving~(\ref{eq:fundamental}), we used the
linear falloff approximation: $K^{(l)}\approx k_1ac^{-l}$.  As~we saw, this
is justified only if $n=ac^{-l} \ll f^{1/2}$.  This will be the case for
the dominant terms in the sum, provided that $\epsilon^{1/2}\ll f^{1/2}$.
So if $\alpha<1$, the formula~(\ref{eq:fundamental}) is valid in the
Kramers limit.  But if $\alpha\ge1$, it does not apply.

In~fact, the $\alpha=1$ case is a crossover regime, in which the ${\cal
O}(f^{1/2})$ length scale is comparable to the ${\cal O}(\epsilon^{1/2})$
length scale.  When $f=f_1\epsilon$ for fixed~$f_1$, the behavior of the
${\cal O}(\epsilon^{1/2})$ boundary layer in the Kramers limit was
determined by Smelyanskiy et al.~\cite{Smelyanskiy99}.  Presumably, their
perturbatively derived expression interpolates between the boundary layer
$h_\epsilon^{(f)}(\bullet,\bullet)$ (as~$f_1\to\infty$) and the $f=0$
boundary layer of Kramers (as~$f_1\to0$).

In closing, we wish to emphasize the experimental importance of the scaling
regimes with $\alpha>0$.  Any system with periodic forcing~$f$ and noise
strength~$\epsilon$ lies on an infinity of curves of the form
$f\propto\epsilon^\alpha$, indexed by~$\alpha$.  It~is the task of the
experimenter to determine which of the corresponding Kramers limit
behaviors, if~any, applies.

This research was supported in part by NSF grant PHY-9800979.

\end{document}